\documentclass[aps,prl,twocolumn,showpacs,groupedaddress]{revtex4}

\usepackage{amsmath}
\usepackage{graphicx}
\usepackage{dcolumn}
\usepackage{bm}

\begin{document}

\title{Bose-Einstein Quantum Statistics and the 
Ground State of Solid $^{\bf 4}$He}

\author{C. Cazorla,$^{{\rm a},{\rm b}}$ G. E. Astrakharchik,$^{\rm c}$ J. 
Casulleras,$^{\rm c}$ and J. Boronat$^{\rm c}$}
\affiliation{
$^{\rm a}$ London Centre for Nanotechnology, UCL, London WC1H OAH, UK \\
$^{\rm b}$ Department of Physics and Astronomy, UCL, London WC1E 6BT, UK\\
$^{\rm c}$ Departament de F\'\i sica i Enginyeria Nuclear, Campus Nord
B4-B5, Universitat Polit\`ecnica de Catalunya, E-08034 Barcelona, Spain}

\date{\today}
\begin{abstract}
The ground state of solid $^4$He is studied using the diffusion Monte Carlo
method and a new trial wave function able to describe the
supersolid. The new wave function is symmetric under the exchange of
particles and reproduces the experimental equation of state.
Results for the one-body density matrix show the existence of off-diagonal
long-range order with a very small condensate fraction $\sim 10^{-4}$. The
superfluid density of the commensurate system is below our resolution 
threshold, $\rho_s/\rho < 10^{-5}$. With a 1\% concentration of vacancies 
the superfluid density is manifestly larger, $\rho_s/\rho=3.2(1) \cdot 10^{-3}$.
\end{abstract}


\maketitle


Solid $^4$He is far from
being a classical crystal as proved by its high degree of anharmonicity,
large kinetic energy per particle ($T \sim 24$ K),
and significant displacements, measured by the Lindemann ratio $\gamma
\sim 0.26$ (to be compared with classical solids, with $\gamma \sim 0.03$
near melting).  
The  counterintuitive possibility of simultaneous solid order and
superfluidity in solid $^4$He has attracted the attention of both theory
and experiment from long time, starting with the theoretical proposal of
Andreev and Lifshitz~\cite{andreev} about a possible supersolid phase related 
to the presence of
a finite concentration of vacancies. Finite values of the superfluid
density and/or condensate fraction in solid $^4$He would be a
macroscopic effect induced by Bose-Einstein statistics. After several
unfruitful experiments to detect superfluid signals in solid $^4$He, Kim
and Chan reported in 2004 the first evidence of non-classical rotational
inertia (NCRI) both in a confined environment~\cite{moses2} and in
bulk~\cite{moses1}. From then on,
several other experimental groups have observed NCRI using different
samples containing small or ultra small $^3$He concentrations, in a simple
crystal or in a polycrystal, and using several annealing
schemes~\cite{balibar_review}. There is an
overall agreement of all the data concerning the onset temperature
$T_0=75-150$ mK at which the superfluid fraction becomes zero, the lowest
value corresponding to ultra pure samples (only 1 ppb $^3$He). Much more
controversial is the value of the superfluid density since the
experimental values reported so far change by more than one order of magnitude ($\rho_s/\rho
\simeq 0.03-0.5$\%) depending on the
purity, annealing, conditions in which the crystal is grown,  etc.   
This high dispersion has led to think that the
superfluid signal observed in solid $^4$He is probably due to the presence of some defects in
the crystal, which could be of different nature: dislocations, vacancies or
grain boundaries. In fact, superfluid flow has been detected when grain
boundaries are present~\cite{sasaki} and also recent measurements of
the specific heat has shown a broad peak at the same onset temperature
$T_0$~\cite{lin}.

Theoretical calculations at very low temperatures, based on the path integral 
Monte Carlo method (PIMC), have not been able to reproduce the
experimental findings on the supersolid. PIMC results show that a
commensurate perfect crystal does not exhibit neither superfluid
fraction~\cite{ceper1} nor
condensate fraction~\cite{ceper2}. Finite values of $\rho_s/\rho$ have been observed only
when disorder is introduced in the form of a glassy phase~\cite{glass} or through
dislocation lines~\cite{dislocation}, but puzzlingly no signal of the superfluid transition is
obtained over the temperature range observed in experiments. 
As the critical temperature of a supersolid is not known, conceptually any
finite-temperature method can not provide the definite answer and one
has to resort to a strictly zero-temperature calculation.
In this letter, we propose a new trial wave
function which allows simultaneously for spatial solid order and Bose symmetry,
and with the benefit of a simple use for importance sampling in diffusion
Monte Carlo (DMC) calculations.

Structure and energetic properties of solid $^4$He at zero temperature have
been widely studied in the past using the Nosanow-Jastrow trial wave
function,
\begin{equation}
\Psi_{\rm{NJ}}({\bf r}_1,\ldots,{\bf r}_N) = \prod_{i<j}^{N} f(r_{ij}) \,
\prod_{i,I=1}^{N} g(r_{iI}) \ ,
\label{njtrial}
\end{equation}  
$N$ being the number of particles and lattice sites (commensurate crystal), 
$f(r)$  a two-body correlation function,
and $g(r)$ a one-body localization factor which links every particle $i$ to
its site $I$. The wave function $ \Psi_{\rm{NJ}}$  leads to an excellent
description of the equation of state and structure properties of the
solid~\cite{cazorla},
but it can not be used for estimating properties which depend directly on
the Bose-Einstein statistics since it is not symmetric under the exchange
of particles. The symmetry requirement can be formally written as
\begin{equation}
\Psi_{\rm{PNJ}}({\bf r}_1,\ldots,{\bf r}_N) = \prod_{i<j}^{N} f(r_{ij}) \,
\left(  \sum_{P(J)} \prod_{i=1}^{N} g(r_{iJ}) \right) \ ,
\label{pnjtrial}
\end{equation}       
with a sum over all the permutations $P(J)$ of lattice sites. This wave
function has been used in the past in variational Monte Carlo (VMC)
calculations by introducing an explicit sampling over the permutation
space~\cite{simetric}.
However, this sampling is inherently inefficient and 
difficult to incorporate in a DMC code. To avoid the complexity of random
walks in the permutation space one can approximate $ \Psi_{\rm{PNJ}}$
by~\cite{zhai}
\begin{equation}
\Psi_{\rm{LNJ}}({\bf r}_1,\ldots,{\bf r}_N) = \prod_{i<j}^{N} f(r_{ij}) \,
 \prod_{i=1}^{N} \left( \sum_{J=1}^{N} g(r_{iJ}) \right) \ .
\label{lpnjtrial}
\end{equation} 
This wave function fulfills the symmetry requirement but it is not well-suited
for genuinely describing solid $^4$He. 
The problem with using $\Psi_{\rm{LNJ}}$ as importance sampling function in DMC
is that the solid phase melts, arriving to liquid (L) or glassy configurations.
This outcome can be understood by analyzing the behavior of
the localization factor (second term in (\ref{lpnjtrial})) under the
per-site occupancy number. Contrary to what is physically reasonable,
multiple occupancy per site is only suppressed via the Jastrow part 
(first term in (\ref{lpnjtrial})). To preserve the solid structure it is
necessary to take into account, already in the localization factor, a
penalization accounting for the voids in the original sites created by
multiple occupancy, a feature missing in the $\Psi_{\rm{LNJ}}$ wave
function.

We introduce a new type of wave function $\Psi_{\rm{SNJ}}$ in which this
feature is present and at the same time the 
necessary requirements of solid order and Bose symmetry are fulfilled. 
The key point is to use the site occupancy as the building block for the
localization factor, thus voids are unequivocally taken into account.
Our model takes the form
\begin{equation}
\Psi_{\rm{SNJ}}({\bf r}_1,\ldots,{\bf r}_N) = \prod_{i<j}^{N} f(r_{ij}) \,
 \prod_{J=1}^{N} \left( \sum_{i=1}^{N} g(r_{iJ}) \right) \ ,
\label{snjtrial}
\end{equation} 
where the product in the second term runs over sites instead of particles,
and the localization factor properly suppresses the voids arising due to
double occupancy, a right behavior which is also
present in the more general wave function $\Psi_{\rm{PNJ}}$.
There have
been other proposals for the study of the supersolid which do not rely on
the symmetrization of the Nosanow-Jastrow model: a Bloch-like
function~\cite{simetric},
inspired in the band theory for electrons, and the shadow wave
function~\cite{shadow}. 
The first model was used in the past in
VMC simulations of Yukawa solids~\cite{simetric} and proved to be difficult to optimize and
provided worse energetic results than the NJ model.
More successful has been the shadow wave
function which has been applied recently in VMC calculations of supersolid
$^4$He~\cite{shadow}. 
However, it has been never applied as importance sampling wave function in a DMC
calculation. Recently, it has been implemented in the path integral ground state
(PIGS) method to eliminate the variational constraints and applied to the
study of two-dimensional solid $^4$He~\cite{shadow2}.
 
The ground state of solid $^4$He has been studied by applying the DMC
method 
to $N$ atoms in a simulation box with periodic boundary conditions. The DMC
method solves stochastically the imaginary-time ($\tau$) Schr\"odinger equation,
providing exact results for boson systems within controllable 
statistical errors.
When $\tau \to \infty$, the ground state dominates and one has a
collection of walkers $\bf{R}_i\equiv\{\bf{r}_1,\ldots,\bf{r}_N\}$ which
follow the probability distribution function ($\Psi_0 \Psi$), $\Psi_0$ and
$\Psi$ being the ground-state wave function and the trial wave function for
importance sampling, respectively. The short-time Green's function
according which the walkers evolve is accurate to order $(\Delta
\tau)^3$~\cite{boro}
and internal parameters of the calculation such as the mean population of
walkers and time step $\Delta \tau$ have been adjusted to eliminate any
bias. 

We have carried out DMC simulations of solid $^4$He using an hcp
lattice with $N=180$ and 448 atoms and a fcc one with $N=108$; the range of
densities analyzed starts slightly below melting, $\rho=0.470\ \sigma^{-3}$
($\sigma=2.556$~\AA) with pressure $P=21$ bar, and ends at $\rho=0.600\
\sigma^{-3}$ with $P=160$ bar. The Jastrow factor in  $\Psi_{\rm{NJ}}$
(\ref{njtrial}) and   $\Psi_{\rm{SNJ}}$ (\ref{snjtrial}) is of McMillan
type, $f(r)= \exp[-0.5 (b/r)^5]$, and the Nosanow term is in both cases a
Gaussian, $g(r)=\exp(-0.5 \beta r^2)$. The parameters $b$ and $\beta$
have been optimized using VMC, the optimal values being $b=1.12\ \sigma$
and $\beta=7.5\ \sigma^{-2}$, and we have neglected their slight density
dependence.

\begin{figure}[tb]
\centerline{
        \includegraphics[width=0.75\linewidth,angle=0]{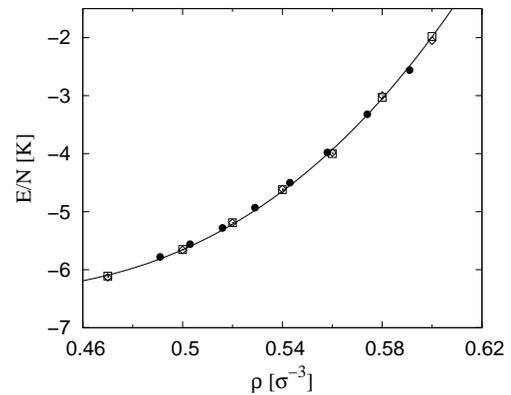}}%
        \caption{DMC results for the energy per particle of solid $^4$He as
	a function of the density. Squares and diamonds stand for the
	symmetric (SNJ)  and non-symmetric (NJ) wave functions, respectively. 
	The solid
	line is a polynomial fit to the SNJ results. Experimental data
	from Ref. \cite{eos-expt} is plotted with solid circles.}
\end{figure} 

In Fig. 1, we show DMC results for the energy per particle $E/N$ as a
function of the density for both the symmetric SNJ (\ref{snjtrial})
and non-symmetric NJ (\ref{njtrial}) wave functions. The calculations used
an hcp lattice with $N=180$ atoms and size corrections to the energy were
estimated using the $N$-dependence observed in VMC calculations of the same
system. In a recent work~\cite{cazorla}, we have verified that this procedure improves the
equation of state with respect to the standard approach of 
assuming that the system is homogeneous beyond $L/2$, with $L$ the length
of the simulation box. As is known, the NJ model is able to reproduce the
experimental equation of state with high accuracy; our present results
confirm this feature as can be observed in Fig. 1. More importantly, the
results obtained with the right symmetrization (SNJ) are statistically
indistinguishable of the NJ ones and are in agreement with
experiment. On the contrary, we have carried out some simulations with the
LNJ model (\ref{lpnjtrial}) and found that the energies are larger than the
SNJ and NJ ones and the solid order is lost, the spatial structure
resembling a glassy state. 

\begin{figure}[tb]
\centerline{
        \includegraphics[width=0.75\linewidth,angle=0]{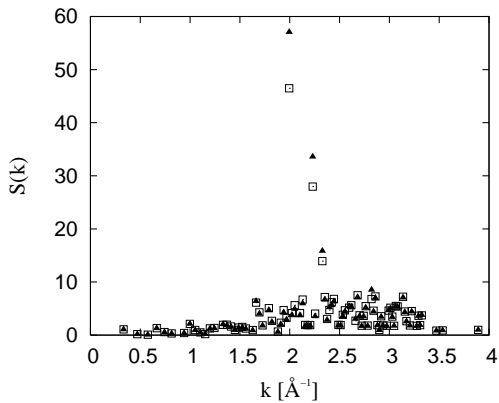}}%
        \caption{Static structure function $S(k)$ at density $\rho~=~0.491\
	\sigma^{-3}$ ($N=180$) calculated with the right symmetry (SNJ) (squares) and
	with the NJ model (triangles).}
\end{figure}

One of the most clear signatures of the solid phase is drawn from the form
of the static structure function $S(k)=1/N \, \langle \rho_{\bf k}
\rho_{-\bf k} \rangle$, with $\rho_{\bf k}= \sum_{i=1}^{N} e^{i {\bf
k}\cdot {\bf r}_i}$. In Fig. 2, we show DMC results of $S(k)$ for hcp
solid $^4$He calculated using the pure estimator method. The results
presented correspond to a density $\rho=0.491\ \sigma^{-3}$ and have been
obtained for the symmetric (SNJ) and non-symmetric (NJ) wave functions.
Both results look rather similar, with the three main singular peaks
corresponding to the reciprocal lattice points clearly visible. The
height of these peaks for the symmetric system is slightly smaller than for
the NJ one but these results confirm that the solid order is preserved by
$\Psi_{\rm{SNJ}}$.   The small decrease in the strength of the main $S(k)$
peaks is probably related to the atomic diffusion we have observed along
the simulations. By monitoring the distance of any particle to its initial
position along the simulation one can know if there 
is atomic diffusion in the crystal as a consequence of particle exchanges.
Our results show that $\sim 15$\% of the atoms present
displacements $r/a > 2$, with $a$ the lattice constant; obviously, the same
measurement for the NJ case does not show any of them. On the other hand, the 
relevance of these exchanges is not observed in the density profile around
the lattice points $\mu(r)$. The results of $\mu(r)$ obtained with  SNJ and NJ are
statistically indistinguishable and therefore both Lindemann ratios
have a common value $\gamma=0.26$, in agreement with experimental data.  

\begin{figure}[tb]
\centerline{
        \includegraphics[width=0.8\linewidth,angle=0]{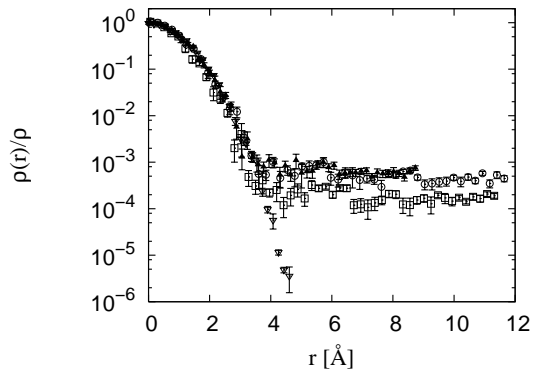}}%
        \caption{One-body density matrix $\rho(r)/\rho$ of solid $^4$He at
	densities $\rho=0.491$ (circles) and $\rho=0.535\ \sigma^{-3}$
	(squares). Full triangles stand for a calculation at $\rho=0.491\
	\sigma^{-3}$ with $N=180$ particles and open triangles for the NJ
	result.}
\end{figure}

A well-known drawback of the NJ model is the impossibility of answering the
fundamental question about the possible existence of off-diagonal long range
order (ODLRO) in solid $^4$He. Instead, the symmetrized model SNJ fulfills
the right Bose-Einstein statistics and therefore is able to provide this
information. Quantitatively, ODLRO is measured by the condensate fraction
$n_0$, that is estimated through the asymptotic behavior of the one-body
density matrix $\rho(r)/\rho$, $n_0= \lim_{r \to \infty} \rho(r)/\rho$. 
Results for $\rho(r)/\rho$ at densities $\rho=0.491$ and $\rho=0.535 \
\sigma^{-3}$, corresponding to pressures $P=31$ and 68 bar respectively,
are reported in Fig. 3. At both densities one can see unambiguously the
existence of ODLRO, and from the asymptotic behavior we estimated the values
of $n_0$: $4.3(2) \cdot 10^{-4}$ and $1.7(4) \cdot 10^{-4}$ at $\rho=0.491$
and $\rho=0.535\ \sigma^{-3}$, respectively (figures within parentheses
correspond to the errors). For comparison, we show in the figure results
for $\rho(r)$ obtained with the NJ model. In order to analyze the influence of the 
number
of particles used in the simulation, we have carried out calculations of
$\rho(r)$ with $N=180$ and 448. The results obtained are plotted in  Fig. 3
at the lower density; within the statistical error both estimations lead to
the same value of $n_0$. All the results plotted in Fig. 3 have been
obtained using extrapolated estimators fulfilling the condition that 
extrapolations of the same accuracy, namely $n_0 \simeq 2 n_0^{\rm mix} - n_0^{\rm var}$
and $n_0 \simeq (n_0^{\rm mix})^2/n_0^{\rm var}$ (mix and var stand for
mixed and variational estimators, respectively), coincide. This equality is
achieved by slight variation of the parameters of the trial wave
function (\ref{snjtrial}) which otherwise does not modify the energy results. 
It is worth mentioning that we have made some VMC calculations removing the
Jastrow part of the SNJ trial wave function and, in all cases, a finite value
of the condensate fraction is observed. This result points to a property 
which seems to be inherent to the symmetrization of the NJ model.

\begin{figure}[tb]
\centerline{
        \includegraphics[width=0.8\linewidth,angle=0]{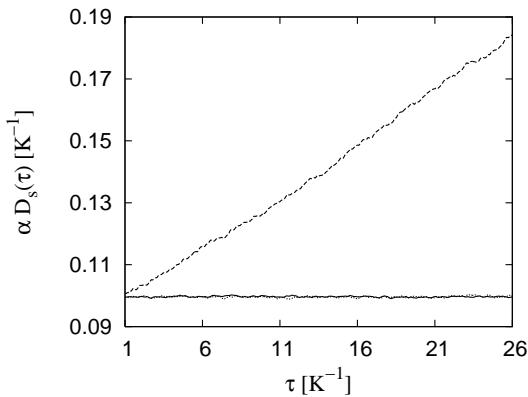}}%
        \caption{Diffusion of the center of mass $\alpha D_s(\tau)$ as a function
	of the imaginary time $\tau$ for solid $^4$He at $\rho=0.491\
	\sigma^{-3}$. The slope of the lines is directly the superfluid
	density $\rho_s/\rho$. The solid line stands for the commensurate
	solid; the dotted line for the case with $N-1$ sites; the dashed
	line for a solid with a $\sim 1$\% concentration of vacancies.}
\end{figure} 

From the different experimental measures carried out on the supersolid, it
seems clear the the presence of vacancies, dislocations or other
defects can be crucial to understand the dispersion on the superfluidity results. 
We have used the SNJ model to determine this influence in the condensate
fraction and superfluid density for two particular cases: a vacancy, i.e.,
a system with $N$ lattice sites and $N-1$ particles, and the absence of a
lattice site, i.e., $N$ particles and $N-1$ sites. Our simulations show
that with $\sim 1$\% vacancy concentration the condensate fraction raises
nearly a factor two with respect to the commensurate case: at $\rho=0.491\
\sigma^{-3}$, $n_0=8.7(4) \cdot 10^{-4}$. On the contrary, the effect of
removing one of the sites in a 1\% fraction does not modify the commensurate value 
$n_0=4.3(2) \cdot 10^{-4}$.

The superfluid density of a bosonic system can be calculated with DMC by
extending the winding-number technique, originally developed for PIMC
calculations, to zero temperature~\cite{zhang}. Explicitly,
$\rho_s/\rho = \lim_{\tau \to \infty} \alpha (D_s(\tau)/\tau )$,
where $\alpha= N/ 6 D_0$ with $D_0= \hbar^2/2m$, and $D_s(\tau)=\langle
({\bf R}_{\rm CM}(\tau) -
{\bf R}_{\rm CM}(0))^2 \rangle$ with $\bf{R}_{\rm CM}$ the center of mass of
the particles in the simulation box. In Fig. 4, results for the function
$\alpha D_s(\tau)$ at a density $\rho=0.491\ \sigma^{-3}$ are plotted as a
function of the imaginary time $\tau$. The slope of this function is
directly the superfluid density $\rho_s/\rho$. Our results show that $\rho_s/\rho$
for the commensurate solid is smaller than our precision limit, that we 
have established at $1 \cdot 10^{-5}$. We have carried out simulations with
different number of particles with the same result. Also we have
calculated $D_s(\tau)$ at higher pressure and at a pressure below melting where 
the solid is metastable obtaining identical conclusion. 
No significant difference with the regular system has been detected for
the case with $N-1$ sites, also shown in Fig. 4. On the other hand, when 
vacancies are present in the crystal with a $\sim 1$\% concentration a
superfluid signal clearly emerges; from the slope, $\rho_s/\rho=
3.2(1) \cdot 10^{-3}$.

To summarize, we have introduced in the microscopic description of solid
$^4$He a new wave function (SNJ) which properly symmetrizes the well-known
Nosanow-Jastrow model. Our results, based on the essentially exact DMC
method, prove that the SNJ trial wave function used for importance sampling
in the method is able to simultaneously reproduce the experimental equation
of state (as the NJ does) and predict results for the condensate fraction 
and superfluidity due to its Bose symmetry. The one-body density matrix shows
ODLRO with a very small condensate fraction,  $n_0=4.3(2) \cdot 10^{-4}$
near melting. A finite but smaller value has been obtained by Galli
\textit{et al.}~\cite{shadow} using VMC with the shadow wave function. On the contrary,
PIMC simulations at finite temperature do not observe ODLRO~\cite{ceper2}. 
An important conclusion of our work is that the zero-temperature
upper-bound for the superfluid fraction $\rho_s/\rho < 1 \cdot
10^{-5}$~\cite{aftalion} is incompatible with 
recent experimental measurements($3\cdot 10^{-4}$-$5\cdot
10^{-3}$)~\cite{balibar_review}, ruling out an explanation of the
superfluid signal as that of a supersolid in a perfect crystal.
The introduction of  1\% vacancies in the
system increases the condensate fraction and a clear signature of
superfluidity is detected. Work is in progress to understand microscopically 
the connection of other defects and/or disorder with both $\rho_s$ and $n_0$.

We acknowledge partial financial support from DGI (Spain) Grant No.
FIS2005-04181 and Generalitat de Catalunya Grant No. 2005SGR-00779.

\end{document}